\documentclass[twocolumn,10pt,prl,aps,showpacs]{revtex4}

\def\3nab{\tilde{\nabla}}

\def\be {\begin{equation}}
\def\ee {\end{equation}}
\def\bea {\begin{eqnarray}}
\def\eea {\end{eqnarray}}

\newcommand{\sfrac}[2]{{\textstyle{#1\over#2}}}
\def\case#1/#2{\textstyle\frac{#1}{#2}}
\usepackage{graphicx}
\usepackage{dcolumn}
\usepackage{bm}
\usepackage{longtable}

\def\cqg{{\em Class. Quantum Grav.\/} }
\def\grg{{\em Gen. Rel. Grav.\/} }
\def\prd{{\em Phys. Rev.\/} {\bf D}}

\def\aph{{\em Ann. Phys. (NY)\/} }
\def\plb{{\em Phys. Lett.\/} {\bf B}}

\begin{document}

\title{Power-law cosmic expansion in $f(R)$ gravity models}

\author{ Naureen Goheer${}^1$, Julien Larena${}^{1}$ and Peter K. S. Dunsby${}^{1,2}$ }

\affiliation{1. Department of Mathematics and Applied Mathematics,
  University of Cape Town, 7701 Rondebosch, Cape Town, South Africa}

\affiliation{2. South African Astronomical Observatory,
  Observatory 7925, Cape Town, South Africa.}

\date{\today}

\begin{abstract}
We show that within the class of $f(R)$ gravity theories,  FLRW  power-law perfect fluid solutions only exist for $R^n$ gravity. This significantly restricts the set of {\it exact} cosmological solutions which have similar properties to what is found in standard General Relativity.  
\end{abstract}
\maketitle
\section{Introduction} 

Currently, one of the most popular alternatives to the {\it Concordance Model} is based on modifications of the
Einstein-Hilbert action.  Such models first became popular in the 1980's because it was shown that they naturally
admit a phase of accelerated expansion which could be associated with an early universe inflationary phase \cite{star80}. 
The fact that the phenomenology of Dark Energy requires the presence of a similar phase (although
only a late time one) has recently revived interest in these models. In particular, the idea that Dark Energy may have a
geometrical origin, i.e., that there is a connection between Dark Energy  and a non-standard behavior of gravitation on
cosmological scales is consequently a very active area of research. 

One such modification is based on gravitational actions which are non-linear in the Ricci curvature $R$ and$/$or contain terms involving combinations of derivatives of $R$ \cite{DEfR,kerner,teyssandier,magnanoff}. Over the past few years, these theories  have provided a number of very interesting results on both cosmological  \cite{ccct-ijmpd,review,cct-jcap,otha,perts} and astrophysical \cite{cct-jcap,cct-mnras} scales. An important feature of these theories is that the field equations can be recast in a way that the higher order corrections are written as an energy\,-\,momentum tensor of geometrical origin describing an ``effective" source term on the right hand side of the standard Einstein field equations  \cite{ccct-ijmpd,review}. In this {\em Curvature Quintessence} scenario, the cosmic acceleration can be shown to result from such a new geometrical contribution to the cosmic energy density budget,  due to higher order corrections of the Hilbert-Einstein Lagrangian.

Of considerable importance to the study of the cosmology of these models is the existence of exact power--law solutions corresponding to phases of cosmic evolution when the energy density is dominated by a perfect fluid. The existence of such solutions is particularly relevant because in Friedmann--Lema\^{\i}tre--Robertson--Walker (FLRW) backgrounds,  they typically represent asymptotic or intermediate states in the full phase--space of the dynamical system representing all possible cosmological evolutions.

In this paper we investigate the implications for the gravitational action if {\it exact} FLRW power--law solutions in $f(R)$ gravity are assumed to exist. We discover that such solutions only occur for a very special class of $f(R)$ theories. This result is complementary to one recently found for $f(G)$ gravity models \cite{goheer09}. 

\section{Field equations for homogeneous and isotropic $f(R)$ models}
We consider the following action within the context of four--dimensional homogeneous and isotropic spacetimes,  
i.e., the (FLRW) universes with negligible  spatial curvature:
\begin{equation}\label{lagr f(R)}
\mathcal{A}=\int d^4 x \sqrt{-g}\left[f(R)+{\cal L}_{m}\right]\;,
\end{equation}
where $R$ is the Ricci scalar, $f$ is general differentiable (at least $C^2$) function of the Ricci scalar and $\mathcal{L}_m$ corresponds to the matter Lagrangian. Units are chosen so that  $c=16\pi G=1$.

It follows that the  field equations for homogeneous and isotropic spacetimes are the {\em Raychaudhuri equation}
\begin{widetext}
\begin{eqnarray}
\dot{\Theta}+\frac{1}{3}\Theta^2=-\frac{1}{2f'}\left[\rho+3P+f-f'R+\Theta f'' \dot{R} +3f'''\dot{R}^2+3f''\ddot{R}\right]]\;,
\end{eqnarray}
\end{widetext}
where $\Theta$ is the volume expansion,  which defines the scale factor $a(t)$ along the fluid flow 
lines via the standard relation $\Theta=3\dot{a}/{a}$, and  $f^{(n)}$ abbreviates 
$\partial^n f/{(\partial R)^n}$ for $n=1..3$;
the {\em Friedmann equation}
\begin{equation}\label{fried}
\Theta^2= \frac{3}{f'}\left[ \rho+\frac{Rf'-f}{2}-\Theta f'' \dot{R}\right]\;;
\end{equation}
the {\em trace equation} 
\be
\label{trace}
3\ddot{R}f''=\rho-3P + f'R-2f-3\Theta f''\dot{R}-3f'''\dot{R}^2\;;
\ee
and the {\em energy conservation equation} for standard matter
\begin{equation}\label{cons:perfect}
\dot{\rho}=-\Theta\left(\rho+P\right)\;.
\end{equation}
Combining the Friedmann and Raychaudhuri equations, we  obtain
\begin{equation}
R=2\dot{\Theta}+\frac{4}{3}\Theta^2\,.\label{R}
\end{equation}
\subsection{Requirements for the existence of power--law solutions}
Analogously to \cite{goheer09}, let us now assume there exists an {\it exact} power--law solution to the field equations,
i.e.,  the scale factor behaves as 
\begin{equation}
a(t)=a_0t^m\;,
\label{a-pl}
\end{equation}
where $m>0$ is a {\it fixed} real number. We further assume that the standard 
matter can be described by a barotropic perfect fluid such that $P=w\rho$ with $w\in[-1,1]$.
From the energy conservation equation, we obtain
\begin{equation}
\rho(t)=\rho_0t^{-3 m (1 + w)}\;,
\end{equation}
and from (\ref{R}) we see that the Ricci scalar becomes
\begin{equation}
R= 6m(2m-1)t^{-2}  \equiv \alpha_m t^{-2}\;.
\label{R_pl}
\end{equation}
Note that $R>0$ if  $m>1/2$, and $R<0$ for $0<m<1/2$, so the value of $m$ fixes the sign of the Ricci scalar.

Using the background solutions above, we can write the Friedmann, Raychaudhuri and trace equations in terms of functions of time $t$ only,  assuming with no loss of generality that $t>0$. 

Considering values of $m\neq 1/2$, we can then solve (\ref{R_pl}) for $t$ and re--write these equations in terms of the Ricci scalar $R$, $f(R)$ and its derivatives with respect to $R$. The Friedmann equation for example becomes
\begin{widetext}
\bea
f''R^2+\frac{m-1}{2}f'R+\frac{1-2m}{2}f+(2m-1)K\left(\frac{R}{\alpha_m}\right)^{\sfrac{3}{2}m(1+w)}=0\;,
\label{friedman-R}
\eea
\end{widetext}
where $K=\rho_0 a_0^{3(1+w)}$. Note that for the power--law solution (\ref{a-pl}), ${R}/{\alpha_m}$ is positive at all times, and therefore equation (\ref{friedman-R}) is real-valued over the range of $R$.

Since we want (\ref{a-pl}) to be a solution at all times, i.e., $R$ spans over an entire branch of the real axis, we can interpret (\ref{friedman-R}) as a differential equation for the function $f$ in  $R$ space.
Solving this equation gives the following general solution
\begin{widetext}
\bea
f(R)=A_{mw}\left(\frac{R}{\alpha_m}\right)^{\sfrac{3}{2} m (1 + w)} + C_1R^{\sfrac{3}{4}-\sfrac{m}{4}+\frac{\sqrt{ \beta_m}}{4}}+\frac{2}{\sqrt{\beta_m}}C_2R^{\sfrac{3}{4}-\sfrac{m}{4}-\frac{\sqrt{ \beta_m}}{4}}\;,
\label{f(R)}
\eea
\end{widetext}
where we have abbreviated
\begin{eqnarray}
A_{mw}=-\frac{4  (2 m-1) \rho_0}{2 - m (13 + 9 w) + 
 3 m^2 (4 + 7 w + 3 w^2)} \;,
 \end{eqnarray}
 \begin{eqnarray}
 \beta_m&=&1+10m+m^2
\end{eqnarray}
and $C_{1,2}$ are arbitrary constants of integration.  We note that the above form 
of $f$ identically satisfies the other field equations, if we similarly convert them into differential equations 
in $R$ space. Note that   $\beta_m>0$ for cosmologically viable solutions with $m>0$. This means that 
the exponents in the solution are all real valued in the case considered here.  Also  $A_{mw}$ is real-
valued and non--zero unless $m=1/2$, but diverges if  $m$ and $w$ satisfy the relationship
$w\equiv\left({3 -7m\pm \sqrt{ \beta_m}}\right)/{6m}$. 
In general, the function $f(R)$ is real--valued if $m$ and $w$ do not satisfy the above relationship, and if $R>0$ (i.e., $ m>1/2$).  Furthermore, If we want to ensure that for  $m=2/[3(1+w)]$ and $K=4/[3(1+w)^2]$ the theory reduces to GR, then we have to set $C_1=C_{2}=0$. In that case, the solution (\ref{f(R)}) is real-valued for all $R$ provided $R/ \alpha_m>0$. If we re-write  $\sfrac{3}{2} m (1 + w)\equiv n$, then we can see that we recover the well-known result that in  $R^n$-gravity, there exists an exact Friedman-like power-law solution $a\propto t^{2n/(3(1+w))}$. The GR--limit can now be identified as the case $n=1$. 
\subsection{Scalar field analogy}
It is interesting to use the solutions found above to reconstruct the effective scalar field often invoked to describe the dynamics of $f(R)$ gravity models. Using this analogy, it has been argued in \cite{Frolov} that $f(R)$ theories suffer from a singularity problem, namely that in the past, at finite time, the dynamics drives the model towards infinite values of the curvature corresponding to points in the scalar field potential atteignable for finite values of the scalar field. Moreover, the effective potential of the models studied in \cite{Frolov} are multivalued, which is a very unnatural feature. In what follows we will show that the models (\ref{f(R)}) that lead to power-law solutions for the scale factor do not suffer from such pathological behaviors, but admit a well-defined scalar field representation with a single-valued potential and no curvature singularity.

The fact that the curvature is well behaved can be directly inferred from Eq. (\ref{R}) since the only divergence occurs for $t=0$, or equivalently, for $a=0$, and this simply corresponds to a standard Big-Bang type singularity.

\begin{center}
\begin{figure}

\includegraphics[width=9cm]{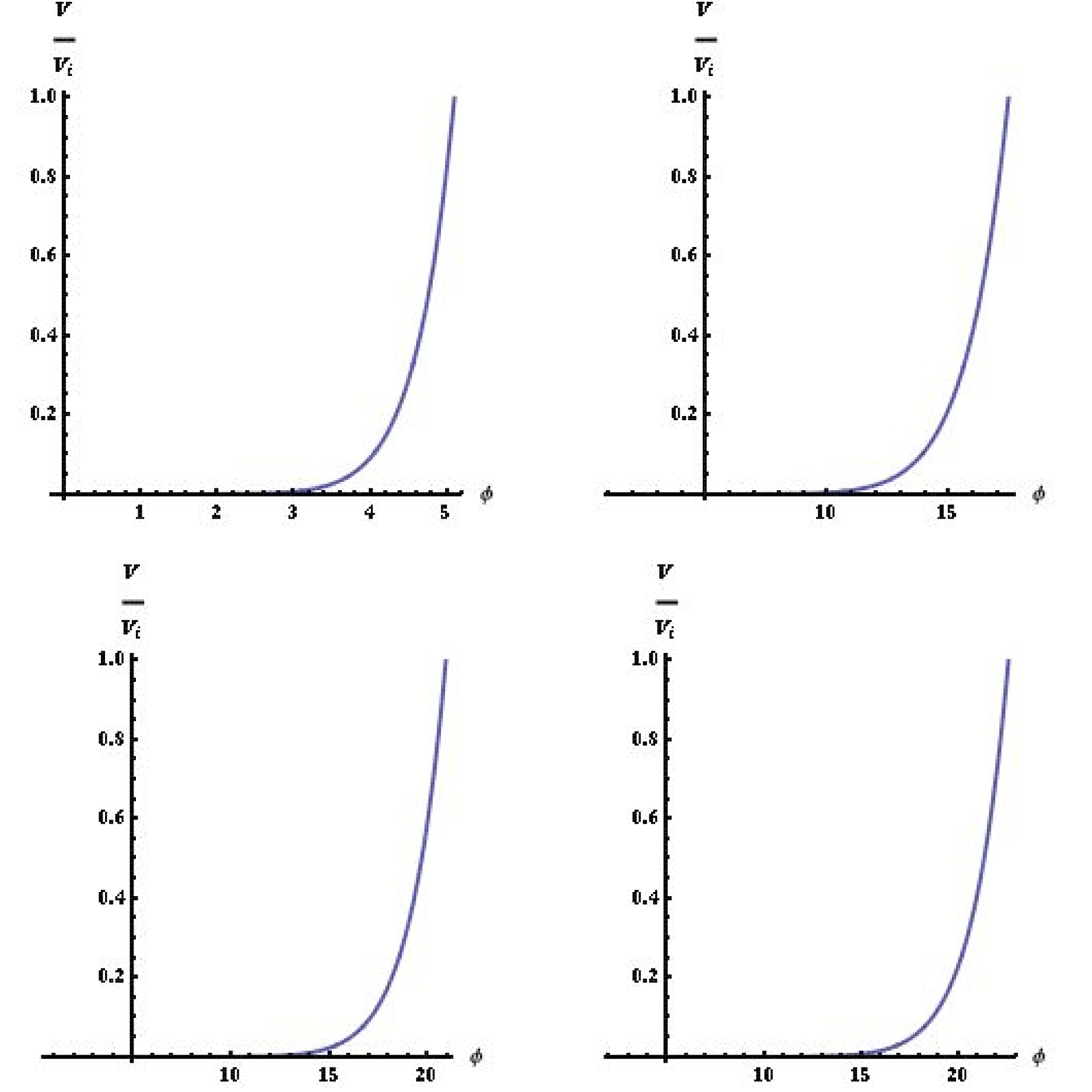}
\caption{Upper left plot: $V(\phi)$ for $w=0$ and $3m/2=1.1$ in the case $C_{1}=C_{2}=0$. This corresponds to $f(R)\propto R^{n}$ with $n=1.1$. Upper right plot: Same model as in the upper left plot but with $C_{1}=C_{2}=\rho_{0}$. Lower left plot: $V(\phi)$ for $w=1/3$ and $3m/2=1.1$ in the case $C_{1}=C_{2}=0$. This corresponds to $f(R)\propto R^{n}$ with $n=1.1$. Lower right plot: Same model as in the lower left plot but with $C_{1}=C_{2}=\rho_{0}$.}
\label{fig1} 
\end{figure}
\end{center}

\begin{center}
\begin{figure}
\includegraphics[width=9cm]{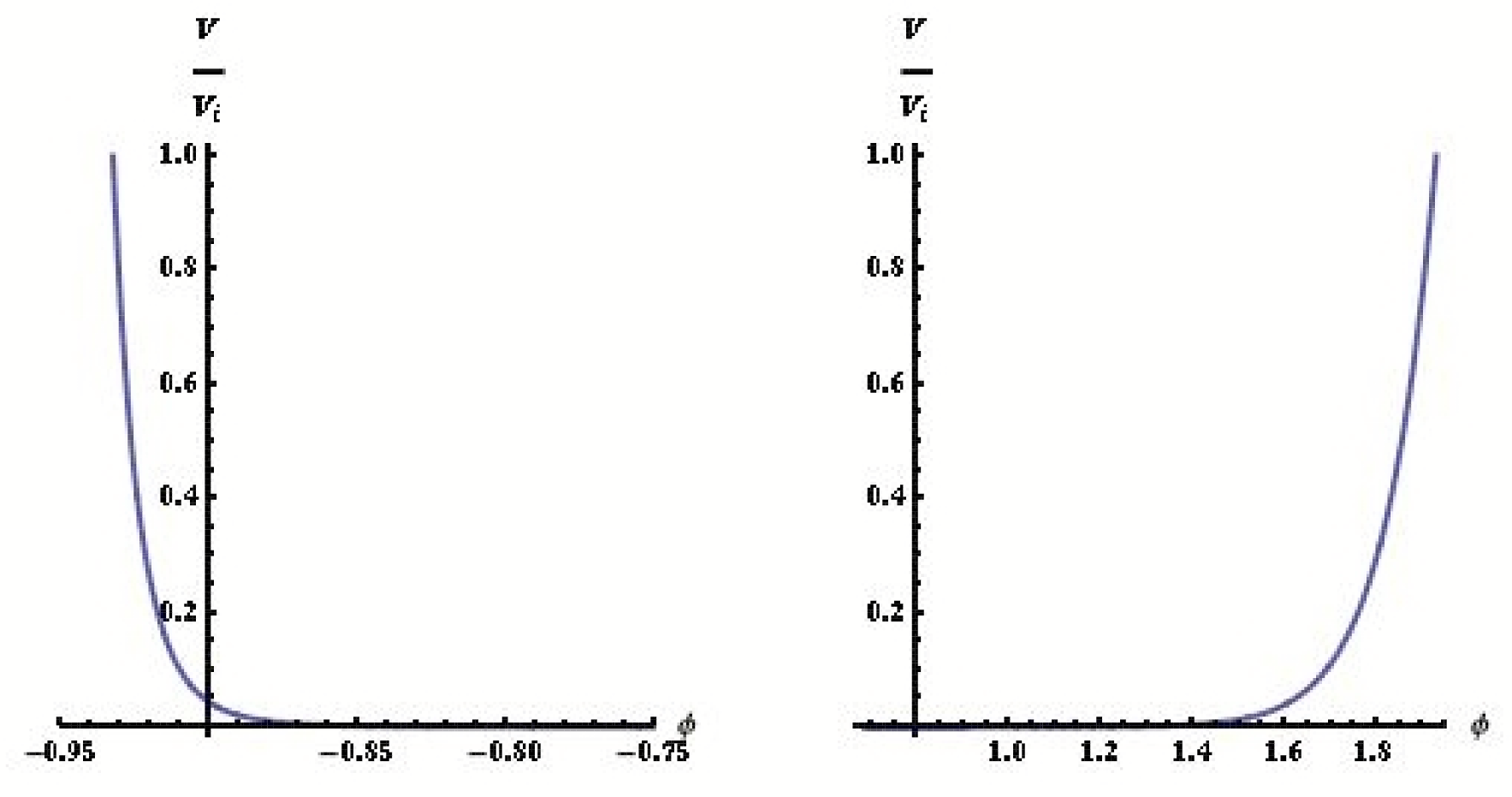}
\caption{Left plot: $V(\phi)$ for $w=0$ and $3m/2=0.9$ in the case $C_{1}=C_{2}=0$. This corresponds to $f(R)$ with the first term in (\ref{f(R)}) proportional to $R^{n}$ with $n=0.9$. Right panel: Same model as in the left plot but with $C_{1}=C_{2}=\rho_{0}$.}
\label{fig2} 
\end{figure}
\end{center}

We adopt the representation in terms of a scalar field used in \cite{Frolov}, by defining the scalar field $\phi$ and its potential $V(\phi)$ through the following equations:
\begin{eqnarray}
\phi&=&\frac{df(R)}{dR}-1\;,\\
\frac{dV}{dR}&=&\frac{1}{3}\left(2f(R)-\frac{df}{dR}R\right)\frac{d^{2}f}{dR^{2}}\;.
\end{eqnarray}
The shape of the potential is illustrated in Figs. \ref{fig1} and \ref{fig2} for various values of the equation of state and of the constants $C_{1}$ and $C_{2}$.

As long as $m>2/3(1+w)$ (which corresponds to $f(R)\propto R^{n}$ with $n>1$ in the case $C_{1}=C_{2}=0$), the characteristic shape of the potential does no depend on the values of $C_{1}$, $C_{2}$, and $w$ ($w$ is in the range of physical values $0<w<1$.). In any case, the scalar field starts at high absolute values and goes down its potential to asymptotically freeze at $\phi=-1$. In the case $m<2/3(1+w)$, the shape of the potential depends on the presence of non-zero $C_{1}$ and $C_{2}$, as illustrated in Fig. \ref{fig2}, but the dynamics nevertheless drives $\phi$ towards a constant value at late times.

\section{Discussion and Conclusion}
We have shown here that \emph{exact} power-law solutions in $f(R)$-gravity can only exist for the very specific form of $f(R)$ given in (\ref{f(R)}). If we ask for the theory to have the correct GR limit, then this $f(R)$ simply reduces to $R^n$. This makes $R^n$ gravity very special in the sense that it is the only $f(R)$ model that allows for exact power-law solutions. Models with actions allowing for terms of the form $R+\tilde{f}(R)$ for example, as studied in \cite{Hu}, do not allow for exact power-law backgrounds unless we reduce to the GR background $\tilde{f}(R)=0$. 

We do not exclude the existence of cosmologically viable trajectories for a general $f(R)$, but suggest that these trajectories may correspond to more complicated exact solutions that asymptotically scale like power-law solutions.

We  emphasize that one must make sure that such background solutions \emph{exist} before performing perturbation theory. It is not enough to simply perturb around an exact power-law background, since this exact background solution does not exist unless $f(R)=R^n$. 

Furthermore,  the above work suggests that in a dynamical systems analysis of any ${f}(R)$ theory other than $R^n$, one should not expect to find any equilibrium points corresponding to exact power-law solutions. 

We conclude that there is a qualitative difference between $R^n$ and any other $f(R)$-model in the sense that $R^n$ has exact power-law solutions even in the non-perturbative non-GR case (e.g. for $n=2$), while any other ${f}(R)$ can only allow these exact solutions in the GR-limit. Therefore, perturbations around these background solutions should be carried out with caution. Moreover, we have shown that the singularity that appears in other $f(R)$ models is not present for the class of models derived from the requirement of a power-law expansion, which is manifest in the scalar field analogy, in which the scalar field potential is a well behaved function of the value of the field $\phi$.

\acknowledgments
The authors would like to thank the National Research Foundation (South Africa) for financial support. JL is supported by the Claude Leon Foundation.


\begin{thebibliography}{99}

 \bibitem{star80}  A. A. Starobinsky, Phys. Lett. {\bf B 91}, 99 (1980);K.~S.~Stelle,
Gen.\ Rel.\ Grav.\  {\bf 9} (1978) 353.

\bibitem{DEfR} Carroll S M, Duvvuri V, Trodden M and Turner M S
2004 \prd {\bf 70} 043528; Nojiri S and Odintsov S D 2003 \prd {\bf
68} 123512; Capozziello S 2002 {\it Int. Journ. Mod. Phys.} D {\bf
11} 483; Mustafa S arXiv: gr-qc/0607116"; Faraoni V 2005 \prd {\bf
72} 124005; Ruggiero M L and Iorio L 2007 JCAP {\bf 0701} 010; de la
Cruz-Dombriz A and Dobado A 2006 \prd {\bf 74} 087501; Poplawski N J
2006 \prd {\bf 74} 084032; Poplawski N J 2007 \cqg {\bf 24} 3013;
Brookfield A W, van de Bruck C and Hall L M H 2006 \prd {\bf 74}
064028; Song, Y, Hu W and Sawicki I 2007 \prd {\bf 75} 044004; Li B,
Chan K and Chu M 2007 \prd {\bf 76} 024002; Jin X, Liu D and Li X
arXiv: astro-ph/0610854; Sotiriou T P and Liberati S 2007  \aph {\bf
322} 935; Sotiriou T P 2006 \cqg {\bf 23} 5117; Bean R, Bernat D,
Pogosian L, Silvestri A and Trodden M 2007 \prd {\bf 75} 064020;
Navarro I and Van Acoleyen K 2007 JCAP {\bf 0702} 022; Bustelo A J
and Barraco D E 2007 \cqg {\bf 24} 2333 Olmo G J 2007 \prd {\bf 75}
023511; Ford J, Giusto S and Saxena A arXiv: hep-th/0612227;
Briscese F, Elizalde E, Nojiri S and Odintsov S D 2007 \plb {\bf
646} 105; Baghram S, Farhang M and Rahvar S  2007 {\bf 75} 044024;
Bazeia D, Carneiro da Cunha B, Menezes R and Petrov A 2007 \plb {\bf
649} 445; Zhang P 2007 \prd {\bf 76} 024007; Li B and Barrow J D
2007 \prd {bf 75} 084010; Rador T arXiv: hep-th/0702081; Rador T
2007 \prd {\bf 75} 064033; Sokolowski L M arXiv: gr-qc/0702097;
Faraoni V 2007 \prd {\bf 75} 067302; Bertolami O, Boehmer C G, Harko
T and Lobo F S N 2007 \prd {\bf 75} 104016; Srivastava S K
arXiv:0706.0410 [hep-th]; Capozziello S, Cardone V F and Troisi A
2006 JCAP {\bf 08} 001; Starobinsky A A arXiv: 0706.2041 [gr-qc]

\bibitem{kerner}  Kerner R 1982 \grg {\bf 14} 453 ;
Duruisseau J P, Kerner R 1986 \cqg {\bf 3} 817.

\bibitem{teyssandier} Teyssandier P 1989 \cqg {\bf 6} 219.

\bibitem{magnanoff} Magnano G, Ferraris M and Francaviglia M 1987 \grg {\bf 19} 465.

\bibitem{ccct-ijmpd} Capozziello S., Cardone V.F., Carloni S.,
Troisi A., 2003, Int.\ J.\ Mod.\ Phys.\  D {\bf 12}, 1969.

\bibitem{review} Capozziello S, Carloni S and Troisi
A 2003 {\it Recent Res. Devel.Astronomy \& Astrophysics} { \bf 1},
625, arXiv: astro-ph/0303041

\bibitem{cct-jcap} S. Capozziello, V.F. Cardone, A. Troisi,
2006, JCAP {\bf 0608}, 001

\bibitem{otha}K.~i.~Maeda and N.~Ohta,
 Phys.\ Lett.\  B {\bf 597} (2004) 400
 arXiv:hep-th/0405205, K.~i.~Maeda and N.~Ohta,
 Phys.\ Rev.\  D {\bf 71} (2005) 063520
 arXiv:hep-th/0411093, N.~Ohta,
 Int.\ J.\ Mod.\ Phys.\  A {\bf 20} (2005) 1
 arXiv:hep-th/0411230,  K.~Akune, K.~i.~Maeda and N.~Ohta,
 Phys.\ Rev.\  D {\bf 73} (2006) 103506
 arXiv:hep-th/0602242.

 \bibitem{perts}
 Kishore N. Ananda, Sante Carloni, Peter K. S. Dunsby,
A characteristic signature of fourth order gravity, arXiv:0812.2028;
Kishore N. Ananda, Sante Carloni, Peter K. S. Dunsby,
A detailed analysis of structure growth in $f(R)$ theories of gravity,
arXiv:0809.3673.

\bibitem{cct-mnras} S. Capozziello, V.F. Cardone, A. Troisi, 2007,
Mon.\ Not.\ Roy.\ Astron.\ Soc.\  {\bf 375}, 1423.

\bibitem{goheer09} N.~Goheer, R. Goswami  P.~Dunsby  \& K. Ananda,  Phys.\ Rev.\ D {\bf 79}, 121301(R) (2009), arXiv:0904.2559.

\bibitem{Frolov}
  A.~V.~Frolov,
  Phys.\ Rev.\ Lett.\  {\bf 101}, 061103 (2008)
  [arXiv:0803.2500 [astro-ph]].

\bibitem{felice08}
A. De Felice, S. Tsujikawa, Construction of cosmologically viable $f(\mathcal{G})$ dark energy models,  arXiv:0810.5712.

\bibitem{Hu} W. Hu  \& I. Sawicki, Phys.\ Rev.\ D {\bf 76}, 064004 (2007), arXiv: 0705.1158;
H. Oyaizu, M. Lima \& W. Hu,  Phys.\ Rev.\ D {\bf 78}, 123524 (2008), arXiv: 0807.2462.


\end{thebibliography}
\end{document}